\documentclass[aps, prb, twocolumn]{revtex4-1}

\usepackage{siunitx}
\usepackage{amsmath, amssymb, amsthm}
\usepackage{graphicx}
\usepackage{caption, subcaption}

\theoremstyle{definition}

\theoremstyle{theorem}

\theoremstyle{remark}

\newcommand{\Ga}{\text{Ga}}

\newcommand{\As}{\text{As}}
\renewcommand{\H}{\text{H}}

\newcommand{\FIGWIDTH}{0.50\textwidth}

\begin{document}

\title{A unified model of droplet epitaxy for compound semiconductor
nanostructures: experiments and theory}

\author{Kristofer Reyes}
\author{Peter Smereka} 
\affiliation{Department of Mathematics, University of Michigan}
\author{Denis Nothern}
\author{Joanna Mirecki Millunchick}
\affiliation{Department of Materials Science and Engineering, University of
Michigan}

\author{Sergio Bietti}
\author{Claudio Somaschini}
\author{Stefano Sanguinetti}
\affiliation{L-NESS, Dipartimento di Scienza dei Materiali, Universita’ di
Milano Bicocca}

\author{Cesare Frigeri}
\affiliation{IMEM-CNR, Parma}

\begin{abstract} 
We present a unified model of compound semiconductor growth
based on kinetic Monte Carlo simulations in tandem with experimental results
that can describe and predict the mechanisms for the formation of various types
of nanostructures observed during droplet epitaxy. The crucial features of the
model include the explicit and independent representation of atoms with
different species and the ability to treat solid and liquid phases
independently.  Using this model, we examine nanostructural evolution in
droplet epitaxy.  The model faithfully captures several of the experimentally
observed structures, including compact islands and nanorings.  Moreover,
simulations show the presence of Ga/GaAs core-shell structures that we validate
experimentally.  A fully analytical model of droplet epitaxy that explains the
relationship between growth conditions and the resulting nanostructures is
presented, yielding key insight into the mechanisms of droplet epitaxy.
\end{abstract}

\maketitle

\section{Introduction}
\label{sec_intro}

Nanotechnology involves the manipulation of matter at the nanoscale in order to
take advantage of the physical properties of materials which arise by a fine
tuning of shape and size. Due to quantum confinement effects, three dimensional
semiconductor nanostructures can behave as artificial atoms and, like natural
atoms, show a discrete spectrum of energy levels \cite{kastner1}.  In contrast
to actual atoms, the electronic properties of quantum nanostructures can be
finely tuned by adjusting structural parameters such as size, composition and
morphology. The latter parameter is the most relevant for the control of quantum
nanostructure electronic properties, as tiny variations in morphology or
composition can cause dramatic changes on the band structure \cite{li1}.

One of the most common methods for the fabrication of quantum nanostructures is
the growth of lattice-mismatched III-V semiconductor materials via the
Stranski-Krastanov (SK) mode. SK growth exploits the self-assembly of
pyramidal-like quantum dots (QDs), driven by the relaxation of lattice
mismatched strain accumulated in the epilayer. Despite the high success of the
technique, which led to fundamental physical understanding and a variety of
applications, the available design degrees of freedom remain limited. The
precise engineering of size and shape of QDs via SK self-assembly remains
problematic, thus limiting the possibilities for the design of arbitrary
electronic and optical properties.

To overcome the SK growth limitations, a fully kinetic limited growth procedure
called droplet epitaxy (DE) was introduced \cite{koguchi1, koguchi3}.  Unlike SK
self-assembly, DE does not rely on strain for the formation of three dimensional
nanostructures. Instead, DE is based on the sequential deposition of III and V
column elements at controlled temperatures and fluxes and, as such, is a
bottom-up fabrication and patterning technique. An initial deposition of
metallic group III such as Ga in a group V free environment leads to the
formation of nanometer size liquid droplets.  Subsequent exposure of the liquid
droplets to the group V vapor causes them to crystallize into quantum dots, by
which we mean compact, three-dimensional nanostructure \cite{koguchi2,
watanabe1}.  A variety of nanostructures may be obtained using this technique,
ranging from compact quantum dots, quantum rings, and core shell structures
\cite{watanabe1, mano1, somaschini1, somaschini2, somaschini3, somaschini4}.
While experimental observations evoke likely mechanisms, a unified model for the
formation of various nanostructural morphologies via DE is still needed.
Understanding the precise mechanisms behind their formation is critical to
controlling the various properties such as shape, size, and composition, which
in turn affect the macroscopic behavior of devices based on such quantum
nanostructures.

Kinetic Monte Carlo (KMC) simulation is a useful tool for developing and testing
models of epitaxial growth and have shed light on the important processes for a
wide range of scenarios.  Previous KMC simulations of GaAs systems modeled
homoepitaxial growth on a cubic lattice. Among the earliest simulations, a
single component was used to model epitaxial growth, varying energy parameters
to account for the expected effect on diffusivity by different Ga/As deposition
ratios \cite{shitara1}. Two-component simulations have also been performed
\cite{kawamura1, kawamura2, ishii1} to study aspects of film growth such as the
transition from two dimensional island formation to step-flow growth.  Such
simulations, however, are necessarily over small areas and are valid only for a
limited range growth parameters.  In most cases these simulations are unable to
model growth regimes in which the group V to group III ratio in the growing film
is not unity, as in the case in droplet epitaxy.

The simulations presented in this paper are designed to capture a wide range of
phenomenology in epitaxial growth, with an emphasis on the growth from the
liquid phase. We chose to examine droplet epitaxy first, but the simulation and
analytical methodologies presented here are applicable to other growth
mechanisms, specifically nanowire grown by the Vapor-Liquid-Solid (VLS) method.
In both VLS and DE growth, a liquid metal reacts with vapor, resulting in
crystallization at the liquid-solid interface. In DE, the metal is consumed
during this crystallization process. In VLS, the metal is either not consumed or
consumed more slowly than in DE, resulting in structures of length scale larger
than the original liquid droplet.

In this paper, we present experimental and simulation results that shed light on
the exact mechanisms behind the formation of nanostructures in droplet epitaxy.
We introduce a KMC model (Section \ref{sec_model}) that is capable of simulating
the formation of nanostructures at all stages, from homoepitaxial growth, to
liquid metal droplet formation and subsequent crystallization.  The model
represents Ga and As atoms explicitly, and treats liquid and solid phases
independently.  Although atomistic, the simulations form large-scale global
structure consistent with experimental results in a reasonable amount of time.
As an initial validation of the model, GaAs homoepitaxy simulations are compared
with analogous experiments, accurately reproducing a surface termination phase
diagram (Section \ref{sec_homoepitaxy}).  We then turn our attention to the main
thrust of the paper, the nanostructural dependence on growth conditions during
droplet epitaxy.  We present experimental results exhibiting this dependence,
along with corresponding simulations (Section \ref{sec_droplet_epitaxy}).  The
model can capture the broad range of nanostructures observed in the experiments
with the correct qualitative dependence on growth parameters.  Moreover,
simulations predict the presence of Ga/GaAs core-shell structures, which are
difficult to observe unambiguously in experiments. The existence of both
monocrystalline and polycrystalline shells suggest two independent mechanisms of
their formation: a morphological instability of the crystallization front or
nucleation at the vapor-liquid interface.  We show by simulation that
nucleation-induced shell structures may be recrystallized into fully
crystallized GaAs islands by annealing in high temperature (Section
\ref{sec_nucleation}).  In the case of the instability-driven shell formation,
we provide simulation and experimental evidence to suggest the presence of a
Mullins-Sekerka instability (Section \ref{sec_ms}).  Lastly, we develop a fully
analytical model (Section \ref{sec_theory}) that describes the existence of the
structures observed and their dependence on growth conditions.  The theoretical
model agrees well with simulation and experimental results.

\section{Kinetic Monte Carlo model}
\label{sec_model}

Prior work in simulating GaAs systems focused on homoepitaxial growth and
studied associated phenomena such as step density \cite{shitara1} and growth
modes of GaAs films\cite{kawamura1}. In these simulations, surface diffusion of
adatoms played the central role, and because of the stoichiometric nature of
epitaxial growth, a simple cubic lattice and the solid-on-solid constraint
sufficed in modeling key aspects of this process. Droplet epitaxy however, poses
several issues that cannot be captured by earlier models. During DE, the system
is inherently non-stoichiometric in that the relative concentrations of Ga and
As atoms on the surface are different from one another and so care must be taken
to model liquid droplet formation.  Processes other than surface diffusion, such
as events within liquid and at the liquid-solid interface, play a key role and
cannot be captured by a simple solid-on-solid model. Lastly, DE results in
large-scale nanostructures on the order of 10s to 100s of nm so that the
simulations must be performed efficiently within large domains.

\begin{figure}[h]
\centering
\includegraphics[width=\FIGWIDTH]{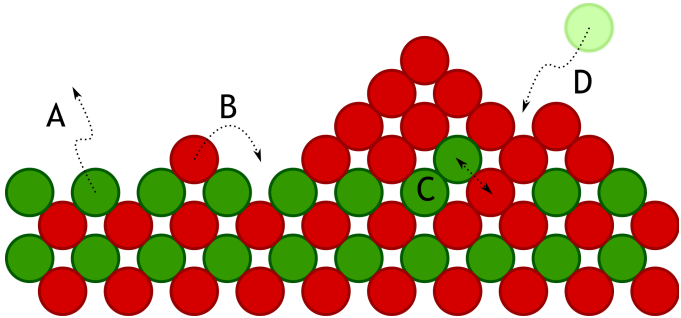}
\caption{
(Color online) KMC Events. Simulations are run on a 1+1 analog of GaAs zinc
blende structure.  Ga atoms are colored red, As atoms are green. The model
allows for atom desorption (A), surface diffusion (B), atom-atom exchanges (C),
and atom deposition (D). The simulations are explicitly multi-species and
atomistic, and evolve via events on individual Ga or As atoms.  
}
\label{fig_events}
\end{figure}

We simulate the homoepitaxial growth and droplet epitaxy experiments with a KMC
model that attempts to address the relevant processes in DE.  In the model,
atoms occupy positions on a 1+1 dimensional analog of the zinc blende lattice
illustrated in Figure \ref{fig_events}.  Each position on the lattice is
adjacent to four nearest neighbors and four next-nearest neighbors. Lattice
sites are vacant or occupied by either a Ga or As atom. The 1+1 dimensional
approach naturally hinders our model to catch effects that are associated to the
three-dimensional crystal structure and to atomic rearrangements driven by
non-local energetics like surface reconstruction phase changes. Specifically
surface reconstruction dynamics leads to changes in adatom surface mobility,
both in terms of diffusion length and diffusivity anisotropy, and in adatom
incorporation into the crystal structure. These effects are at the origin, in
DE, of the observed dependence on substrate temperature and preparation
procedure in the droplet formation critical coverage during the initial Ga
deposition  and of the shape and anisotropy of control of the quantum
nanostructures that are possible thanks to fine control of surface
reconstruction dynamics {\cite{somaschini6, somaschini7}}. The simplicity of the
model implies such effects cannot be captured, in line with previous work on KMC
modeling. We note here that other theoretical models can capture surface
reconstructions {\cite{chakrabarti1}}. The processes that are captured and
studied in this paper, however, yield first-order insight to the growth
mechanisms in DE.

In each Monte Carlo step, one of four local events alters at most two positions
on the lattice.  Atoms may desorb from the surface (A in Figure
\ref{fig_events}), diffuse on the surface (B), exchange with neighboring atoms
(C), or adsorb onto a vacant position on the surface (D). The model does not
enforce a solid-on-solid constraint.  Instead, atom configurations are required
to be connected, i.e. there must exist a chain of atoms (through nearest
neighbor bonds) between any two atoms in the configuration.  While connectedness
is inherently a global property, the requirement is approximated by enforcing
the property on all local neighborhoods. Any event resulting in a disconnected
configuration is disallowed. 

Event rates are determined by simple nearest and next-nearest neighbor
bond-counting. Due to the multi-species nature of the model, several bonding
energies must be specified. Specifically, nearest and next-nearest neighbors are
bonded by energies that depend solely on the species of the two connected atoms.
Nearest neighbor bonding energies are given by three parameters $\gamma(\Ga,
\Ga), \gamma(\Ga, \As)$ and $\gamma(\As, \As)$. Next-nearest neighbor bonds are
only assigned between atoms of the same species and are denoted
$\gamma_{nn}(\Ga, \Ga)$ and $\gamma_{nn}(\As, \As)$. In this way, Ga and As only
interact through nearest-neighbor bonds, whereas Ga-Ga and As-As interact
through nearest and next-nearest neighbor bonds. The parameter values used in
the simulations are summarized in Table \ref{table_energy_params}.  These energy
values imply that the Ga-As bonds are energetically preferred, while weak As-As
bonds effectively eliminate excess As by desorption. The bonding energies were
calibrated from the homoepitaxy (Section \ref{sec_homoepitaxy}) and liquid
droplet simulations (Section \ref{sec_droplet_epitaxy}) to match qualitative
experimental assumptions and observations.  Within the homoepitaxy simulations,
the Ga-As bond strengths were tuned in order to observe a transition from rough
island formation in the low temperature regime to a smooth step-flow growth
mechanism in high temperatures \cite{orme1}. In the case of liquid droplet
simulations, we tune the Ga-Ga bond strength to match liquid droplet statistics
such as droplet width and number density.

\begin{table}[h] \centering
    \begin{tabular}{cc}{
        \begin{tabular}{c|cc}
            $\gamma (\text{eV})$ & Ga  & As  \\ \hline
                              Ga & 0.3 & 0.5 \\
                              As & 0.5 & 0.1 
        \end{tabular}
    } & {
        \begin{tabular}{c|cc}
            $\gamma_{nn} (\text{eV})$ & Ga  & As \\ \hline
                                   Ga & 0.3 & -  \\
                                   As & -   & 0.1 
        \end{tabular}
    } \end{tabular}

\caption{
(Color online) Pairwise nearest and next-nearest neighbor bonding energies used
throughout the paper.
} 
\label{table_energy_params} 
\end{table}

Monte Carlo transition rates $r(X,Y)$ from atom configuration $X$ to
configuration $Y$ are taken to be of an Arrhenius form: 
$$ r(X,Y) = R_0 e^{ -E_a(X,Y) / k_BT },$$ 
where $R_0 = 10^{13} \text{s}^{-1}$ is a constant rate prefactor, $T$ is
temperature, and $E_a(X,Y)$ is the activation energy for a particular event.

For surface diffusion,  $E_a$ is simply the sum of the bonding energies about
the diffusing atom. For the desorption of an As atom, the activation energy is
the sum of the bonding energies about the desorbing atom and an additional
desorption barrier $\mu_{As} = 1.1$ eV, its value calibrated by homoepitaxial
simulations described in Section \ref{sec_homoepitaxy}.  The desorption of Ga is
disallowed.  In the case of atom-atom exchanges, which occur within a liquid
droplet, we use a base energy barrier $\epsilon_D$ for the As diffusion barrier
in liquid Ga, described in more detail below.  To facilitate exchanges occurring
at the liquid-solid interface, for which we must take solid bonds into account,
we define an intermediate state in which the exchanging atoms are replaced by an
intermediate species $\H$. Bonding energies between $\Ga$ or $\As$ atoms and
$\H$ atoms are obtained by averaging over $\H = \Ga$ and $\H = \As$, e.g.  $$
\gamma(\Ga, \H) = \frac{1}{2}\left(\gamma(\Ga, \Ga) + \gamma(\Ga, \As)\right).
$$ The activation energy $E_a$ of an atom-atom exchange at the liquid-solid
interface is then taken to be the sum of the base barrier $\epsilon_D$ and the
change in bonding energies between the original and the intermediate states.

\begin{figure}[h]
\centering
\includegraphics[width=\FIGWIDTH]{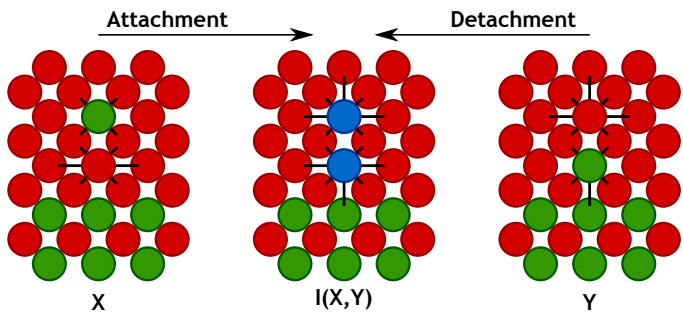}
\caption{
(Color online) Attachment and detachment events at the liquid-solid interface.
This figure illustrates the attachment and detachment of an As atom in liquid Ga
onto and from a perfectly flat liquid-solid interface, along with the
intermediate state for the transitions. The intermediate species $\H$ is colored
blue. The black lines indicate the relevant bonds that contribute to the change
in energy between the initial and intermediate state.
}
\label{fig_attachment_detachment}
\end{figure}

For example, we consider the attachment and detachment of an As atom onto the
liquid-solid interface, as illustrated in Figure
\ref{fig_attachment_detachment}. We denote the states before and after the
detachment as $X, Y$, respectively, and the intermediate state as defined above
by $I(X,Y)$. The change in energy $E(X) - E(I(X,Y))$ between states $X$ and
$I(X,Y)$, describing the additional barrier for attachment can be computed from
a small number of relevant bonds, indicated by black lines in the figure. The
total activation energy for attachment $E_A = \epsilon_D + E(X)-E(I(X,Y))$ is
then given by
\begin{eqnarray*}
E_A 
&=& \epsilon_D + 4\gamma(\Ga, \As) + 4\gamma(\Ga, \Ga) + 2\gamma_{nn}(\Ga, \Ga) \\
& & -8\gamma(\Ga, \H) - 5\gamma_{nn}(\Ga, \H) - \gamma_{nn}(\As, \H),  \\
&=& \epsilon_D  - \frac{1}{2}\gamma_{nn}(\Ga, \Ga) - \frac{1}{2}\gamma_{nn}(\As, \As),\\
&=& 0.5 \text{ eV}.
\end{eqnarray*}
We may calculate the  activation energy for detachment $E_D = \epsilon_D + E(Y)
- E(I(X,Y))$ in a similar manner:
\begin{eqnarray*}
E_D 
&=& \epsilon_D + 4\gamma(\Ga, \As) + 4\gamma(\Ga, \Ga) \\
& & + 3\gamma_{nn}(\Ga, \Ga) + \gamma_{nn}(\As, \As) \\
& & - 8\gamma(\Ga, \H) - 5\gamma_{nn}(\Ga, \H) - \gamma_{nn}(\As, \H), \\
&=& \epsilon_D  + \frac{1}{2}\gamma_{nn}(\Ga, \Ga) + \frac{1}{2}\gamma_{nn}(\As, \As),\\
&=& 0.9 \text{ eV}.
\end{eqnarray*}
The forms of the attachment and detachment barriers $E_A$ and $E_D$ indicate
what physically occurs during the transitions, namely the formation/removal of
one next-nearest neighbor Ga-Ga bond and one next-nearest neighbor As-As bond.

\subsection{Liquid Barriers}

Because we wish to simulate liquid droplet epitaxy, our model identifies liquid
neighborhoods in which events are assigned different energy barriers to account
for the salient physics within a liquid droplet.  The diffusion of As through a
liquid droplet occurs atomistically as an exchange event between the diffusing
As atom and a Ga atom neighbor. Such an atom-atom exchange event is deemed a
liquid event if there is at most one As atom total within the neighborhood of
the exchanging atoms. 

Further, such exchanges in the liquid are categorized as occurring on either the
surface or bulk of a droplet. A surface liquid exchange event is one in where
there is at least one vacancy within the neighborhoods of the exchanging atoms.
These events represent the incorporation of an As atom from the surface into the
bulk of the droplet or \emph{vice versa} and are given a constant activation
energy barrier $\epsilon_{I} = 1.05$ eV. The particular value of $\epsilon_I$ is
slightly larger than the 1 eV barrier for the diffusion of an As adatom on the
surface of a liquid droplet. A liquid event that does not occur on the surface
is categorized as a liquid bulk event and is assigned  the energy barrier
$\epsilon_D$, which represents the diffusion barrier of an As atom within the
bulk of the liquid. The value $\epsilon_D = 0.7$ eV was calibrated by the
droplet epitaxy simulations described in Section \ref{sec_droplet_epitaxy}.
Only after this model was completed did we become aware of the experimental work
of Gorokhov {\it et al.} \cite{gorokhov1} who reported the energy barrier for
diffusion As in Ga of approximately 0.7 eV.

Because of its ability to handle the liquid phase independently, along with its
multi-component nature, the model may be minimally adapted to simulate other
growth modes in which liquid droplets are used. In particular, the model has
been applied to simulate nanowire growth by the VLS method {\cite{reyes1}}.
Here, identifying liquid neighborhoods is important to prescribe special energy
barriers for the conversion of vapor phase to a solid one, mimicking the
catalytic effect of the liquid droplet. Additionally, without the SOS
constraint, the model captures phenomena such as nanowire bending.

\subsection{Detailed Balance}

In order to discuss detailed balance, we consider the behavior of our model in
the absence of deposition and desorption \cite{biehl1}. Under such conditions,
we assume the incorporation of an As atom from the surface of a droplet into the
liquid bulk to be a rare event.  The rates of the remaining events (i.e.
surface diffusion and atom-atom exchanges both within the liquid and at the
liquid-solid interface) satisfy detailed balance with respect to the Boltzmann
distribution 
$$ \pi(X) = \frac{1}{Z} \exp\left[-\frac{E(X)}{k_B T}\right], $$
where $E(X)$ is the energy of the state obtained by bond counting, using bonding
energies in Table \ref{table_energy_params}. For these events $X \rightarrow Y$,
the activation energies are of the form
\begin{equation}
\label{eqn_activation_general}
E_a = E(X) - E(I(X,Y)) + \epsilon(X,Y),
\end{equation}
where $I(X,Y)$ is a transition state between $X$ and $Y$ and $\epsilon(X,Y)$ is
an additional barrier. Detailed balance follows since we have required that
$I(X,Y)$ and $\epsilon(X,Y)$ be symmetric with respect to their arguments.

For surface diffusion, $I(X,Y)$ is the state with the diffusing atom removed and
$\epsilon(X,Y) = 0$. Because the reverse event $Y \rightarrow X$ is also a
surface diffusion event, it follows that $I(X,Y) = I(Y,X)$ and $\epsilon(X,Y) =
0 = \epsilon(Y,X)$.  For atom-atom exchanges, $I(X,Y)$ is the state where the
exchanging atoms are replaced with intermediate species $\H$, and hence $I(X,Y)
= I(Y,X)$, and $\epsilon(X,Y) = \epsilon_D$. Note that it was claimed above that
for an As atom diffusing through liquid Ga, an activation barrier $\epsilon_D$
was used without any further contribution to the barrier from bond counting.
This is consistent with the generalized description of the activation energy in
equation \eqref{eqn_activation_general}, because in this case $E(X) = E(I(X,Y))
= E(Y)$, i.e. no bonds are formed/removed during the diffusion of As in liquid
Ga.

\subsection{Implementation}

Because of the time and length scales associated to the nanostructures we wish
to simulate, an efficient implementation of the KMC algorithm is important. A
rejection-free sampling of rates is achieved by the standard binary-tree data
structure containing partial sums of the rates\cite{blue1}, achieving $\mathcal
O(\log N)$ sampling by binary search and $\mathcal O(\log N)$ tree updates,
where $N$ is the number of atoms. As the algorithm proceeds, event rates for
several atoms must be recomputed at every step. A rate caching technique is
employed to remove repeated rate calculations by taking advantage of the
recurrent nature of local neighborhoods within KMC \cite{reyes1}. This procedure
leads to significant performance gains.

\section{Homoepitaxy}
\label{sec_homoepitaxy}

As an initial application and validation of the simulations, we model GaAs
substrate growth at various growth conditions. It is known that the surface
reconstruction of a GaAs substrate depends on both the temperature and the
relative deposition rates of Ga and As \cite{ohtake1}.  In 1+1 dimensions,
simulations cannot reproduce surface reconstructions. Instead, we measured
surface termination and its dependence on temperature and incoming deposition
rates. The simulation results were then compared to experimental data. 

Experiments on  GaAs(001) were carried out using molecular beam epitaxy (MBE)
equipped with a solid source for Ga and a valved source for As$_4$.  Substrates
were initially heated under As$_4$ overpressure to desorb the native oxide
layer.  A 75 nm GaAs buffer layer was then grown at a substrate temperature of
\SI{590}{\celsius} with a Ga deposition rate of  0.4 monolayers per second
(ML/s) and an As$_4$ growth rate of at least 2 ML/s, as measured by
reflection high energy electron diffraction (RHEED) oscillations
\cite{ferguson1}.  After growing the GaAs buffer, the substrate temperature was
fixed between \SI{460}{\celsius} and \SI{610}{\celsius}. GaAs was then grown
with a Ga rate of 0.6 ML/s while the As rate was slowly reduced from 3.9
ML/s to 1.0 ML/s.  During the growth, the RHEED pattern was monitored for
the change from an As-terminated reconstruction to a Ga-terminated
reconstruction.  Consistent with the expected result, at  high substrate
temperatures the As rate required to maintain an As-terminated surface increases
with temperature. At lower temperatures, the necessary As rate becomes invariant
with temperature.

To simulate GaAs substrate growth, both Ga and As were deposited simultaneously
on initially flat, As-terminated substrate. The Ga deposition rate $F_{Ga}$ was
fixed at 0.37 ML/s, while the As deposition rate $F_{As}$ was varied so that
the deposition ratio $F_{As}/F_{Ga}$ ranged between .5 and 10. Rates (reported
in ML/s)  here and throughout this paper describe the rate at which atoms are
added to the system. The observed stoichiometric growth of the film is an
emergent property of the model rather than explicitly enforced through
deposition rates. This is manifest in the stoichiometry observed over a broad
range of deposition ratios.

\begin{figure}[h]
\centering
\includegraphics[width=\FIGWIDTH]{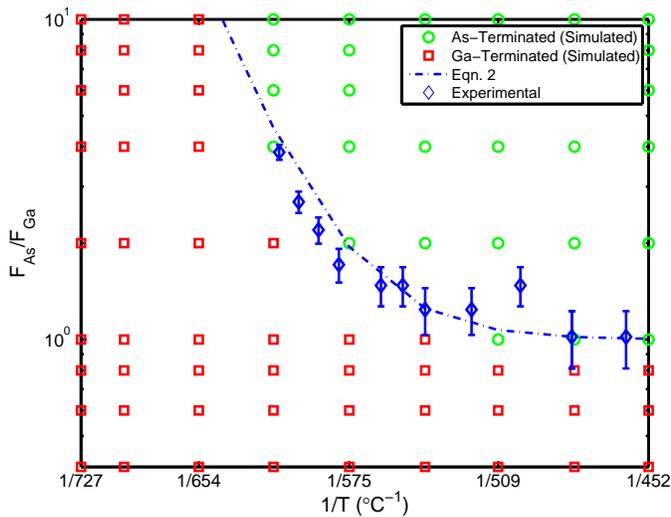}
\caption{ 
(Color online) Substrate termination phase diagram as a function of deposition
ratio and temperature, obtained from simulations and experiments.  Those growth
conditions resulting in a mostly Ga-terminated substrate are in indicated by red
squares, while green circles label As-terminated ones.  The blue points above
indicate the conditions where the transition from Ga to As termination occurred
experimentally. The blue curve indicates the boundary between Ga and
As-terminated given by equation \ref{eqn_surface_termination_boundary}.  
}
\label{fig_phase_diagram}
\end{figure}

The temperature was varied between $\SI{427}{\celsius}$ and
$\SI{727}{\celsius}$.  Five monolayers of total material was deposited, and
surface Ga concentration was measured at regular intervals during the deposition
of the last two monolayers.  Figure \ref{fig_phase_diagram} is a surface
termination phase diagram as a function of deposition rate and temperature for
both experimental and simulation results. Red squares indicate the conditions
where simulations show a predominantly Ga-terminated surface, while green
circles are conditions yielding a predominantly As-terminated surface.
Experimentally determined transition from the As-terminated $2\times 4$ to the
Ga-terminated $4\times 2$ reconstructions as determined by RHEED are shown as
blue diamonds and indicate a good agreement between simulation and experimental
data. 

The relevant parameter that controls surface termination is the As desorption
barrier $\mu_{As}$. Varying this parameter effectively shifts the above phase
diagram horizontally. The value $\mu_{As} = 1.1$ eV is fitted to experiments.
This value, combined with the energy values in Table \ref{table_energy_params}
yield a total activation energy of 
$$2\gamma(Ga, As) + \gamma_{nn}(As,As) + \mu_{As} = 2.2 \text{ eV},$$
for the desorption of an As adatom from a Ga terminated substrate, comparing
favorably to experimental results\cite{liang1, yamaguchi1}. Moreover, the
specific parameter value does not significantly impact the qualitative shape of
the phase diagram.  That is, independent of $\mu_{As}$, the simulations capture
a constant critical deposition ratio in the low temperature regime and its
transition to an increasing critical ratio as temperature increases. 

Due to the simplicity of the model, individual As atoms are deposited on the
surface rather than As$_4$ molecules. In this homoepitaxy study, the
adsorption/desorption processes, and in particular, the value of $\mu_{As}$ was
tuned to match the macroscopic properties surface termination and concentration.
As such, the fitted value of $\mu_{As}$ serves as an accurate measure of As
growth and incorporation onto the substrate.  Other surface As kinetic effects,
including those determined by surface reconstruction dynamics on the local
scale, may not be captured by the model.  However, in the DE growth procedure,
these effects appear to be relevant mostly in determining the extended
nanostructure shape control peculiar to DE {\cite{somaschini6}} or in the
extremely low temperature and high As fluxes. We therefore believe that such
effects should lead to minor modifications in the description of the DE
processes that are studied in this paper.

An analytical expression for the boundary between Ga-terminated and
As-terminated substrates can be determined.  During homoepitaxial growth,
several processes occur, including surface diffusion and As desorption, but to
first order we may approximate the system in a quasi-static deposition/
desorption-limited regime. In this regime, the transition between the As and
Ga-terminated surface occurs when the amount of Ga on the impinging upon the
surface (given by $F_{Ga}$) is equal to the net rate of As growth (given by
$F_{As} - R_{desorb}$), where $R_{desorb}$ is the desorption rate of $As$, taken
as the harmonic average of the desorption rates for an As adatom on a
Ga-terminated surface (with desorption energy barrier 2.2 eV) and a As atom on a
Ga-terminated surface with one nearest or next nearest neighbor (with desorption
barrier 2.3 eV). This boundary is given by the equation 
\begin{equation}
    \frac{F_{As}}{F_{Ga}} = 1 + \frac{R_{desorb}}{F_{Ga}},
    \label{eqn_surface_termination_boundary}
\end{equation}
and agrees well with both simulation and experiments.
       
\section{Liquid Droplet Epitaxy}
\label{sec_droplet_epitaxy}

The main thrust of this work was to accurately capture the processes relevant to
liquid droplet formation and crystallization.  The experiment and simulations
proceed in two main steps, which are at the basis of the droplet fabrication of
GaAs/AlGaAs quantum nanostructures \cite{watanabe1}. First, Ga is deposited onto
the substrate by MBE, forming liquid Ga droplets. Next the droplets are
crystallized by the introduction of an As flux.  Depending on growth conditions
such as temperature and As overpressure, the actual morphology upon
crystallization can vary \cite{watanabe1}. The observed nanostructures may range
from fully crystallized compact quantum dots, to nanorings and even etched holes
\cite{wang1, somaschini1, somaschini6, mano1, kuroda1}. 

\subsection{Quantum Dots and Nanorings} 

\subsubsection{Experimental methodology}
In order to experimentally assess the exact dependence of droplet epitaxy
nanostructures on the growth condition, we prepared a matrix of samples,
sampling the typical droplet epitaxy ranges for substrate temperatures and
As$_4$ beam equivalent pressure (BEP). The samples were prepared following the
standard procedure for droplet epitaxy \cite{watanabe1,  watanabe2, kuroda2}.
After the oxide removal and the GaAs buffer layer growth (\SI{1}{\micro\meter}
on thick at \SI{580}{\celsius}) to ensure the atomic smoothness of the surface,
a 200 nm thick Al$_{0.3}$Ga$_{0.7}$As barrier layer is grown at the same
temperature of \SI{580}{\celsius}. The presence of an AlGaAs barrier permits the
use of the fabricated structures, once capped with another AlGaAs layer, as
morphology controlled quantum nanostructures. We do not expect the Al content of
the barrier to influence the obtained GaAs nanostructure morphology as the
AlGaAs barrier is buried under a 1.75 ML GaAs top layer which forms on the
surface during the following Ga deposition step  \cite{watanabe1, sanguinetti1}.

\begin{figure*}[!htb]
\centering
\includegraphics[width=0.8\textwidth]{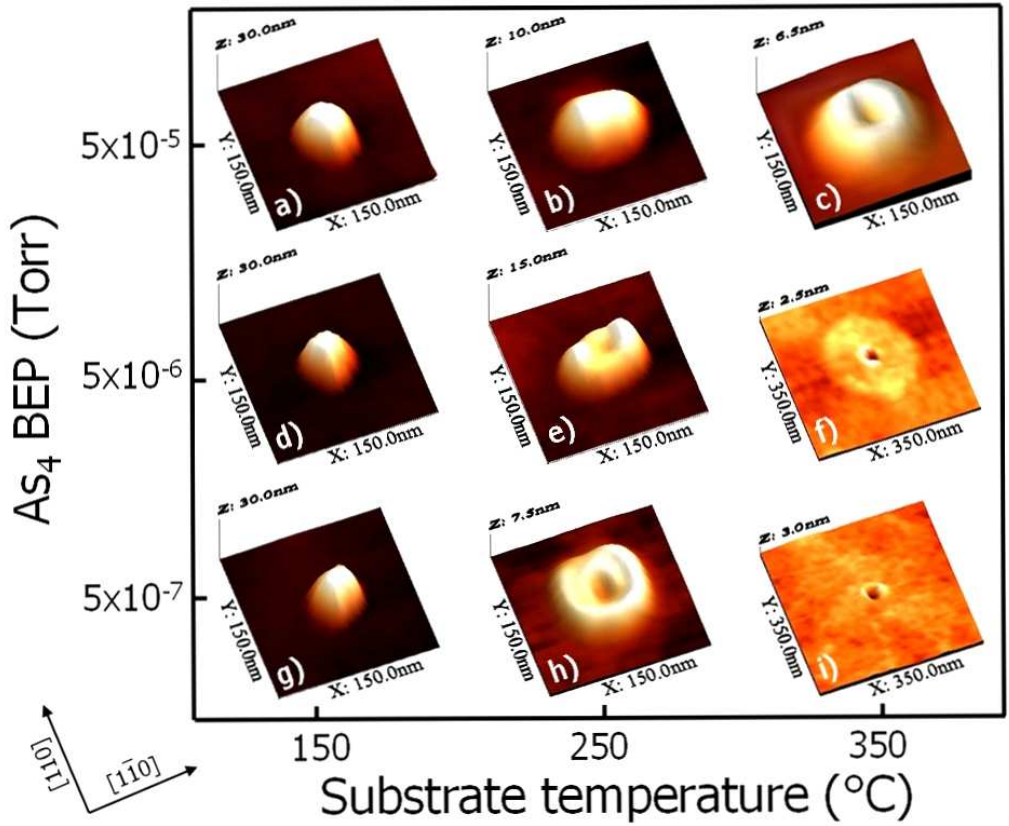}
\caption{
(Color online) Droplet epitaxy experimental results showing typical nanostructures observed
over a range of substrate temperatures and As$_4$ BEP.  
}
\label{fig_zoology}
\end{figure*}

Prior to the deposition of Ga, the As cell is closed and the background pressure
reduced below $10^{-9}$ Torr. Ga droplets are then formed on the substrate
surface by supplying 2.5 ML of Ga at \SI{350}{\celsius} with a deposition rate
of 0.08 ML/s. Many nearly hemispherical Ga droplets form at a density of
$8\times 10^8$ cm$^{-2}$.  The average droplet diameter and height are 50 nm and
20 nm, respectively. The same Ga droplet preparation procedure is used for each
sample in order to assure the same droplet density before the crystallization
procedure. We then explored the nanostructure fabrication parameter space by
varying the substrate temperature and the As$_4$ BEP used for the
crystallization of Ga droplets. Nine different samples are prepared by
systematically varying the temperature between 150 and $\SI{350}{\celsius}$  and
the As$_4$ BEP between $5\times 10^{-7}$ and $5\times10^{-5}$ Torr. The
crystallization time is kept constant (10 min.) to ensure the complete reaction
of the Ga contained in the droplets with As atoms.  The morphological dependence
of the fabricated nanostructures, measured by Atomic Force Microscopy (AFM) on
temperature and As overpressure is summarized in Figure \ref{fig_zoology}. From
this data, it is clear that compact islands form at low temperature, and as the
As overpressure is reduced or the temperature is increased, the nanostructures
become rings.  At the highest temperature and lowest As overpressure, only holes
remain in the place of the droplet.  From large area scans (not shown here) the
number of GaAs nanostructures per unit area are in excellent agreement with the
original droplets density.  Each droplet was thus transformed into a GaAs
nanostructure at the end of the crystallization procedure.

\subsubsection{Simulations}
The simulations were performed in a similar manner to the experiments outlined
above. However, in order to simplify the analysis, the temperature was
maintained constant throughout each simulation.  This is justified by the
observation that the Ga deposition temperature affects droplet density but does
not play a relevant role in determining droplet crystallization dynamics.
Simulations proceed as follows:

\begin{enumerate}
    \item Ga atoms are deposited on a flat, As-terminated GaAs substrate at a
    rate of 0.1 ML/s at temperature $T$ until 4.0 monolayers are deposited; 
    \item The system is then annealed for 60 seconds in the absence of
    deposition; 
    \item After annealing, an As flux is introduced by the deposition of As at a
    rate of $F_{As}$ ML/s until the system attains equilibrium.  
\end{enumerate}

The growth parameters $T$ and $F_{As}$ were varied in order to study their
effect on the resulting morphology. The temperature $T$ ranged between
\SI{150}{\celsius} and \SI{350}{\celsius}, while the As deposition rate $F_{As}$
ranged between 0.1 ML/s and 4 ML/s Within this range of growth parameters,
we are able to simulate the formation of a variety of nanostructures similar to
those observed experimentally.

\begin{figure}[!htb]
\centering
\includegraphics[width=\FIGWIDTH]{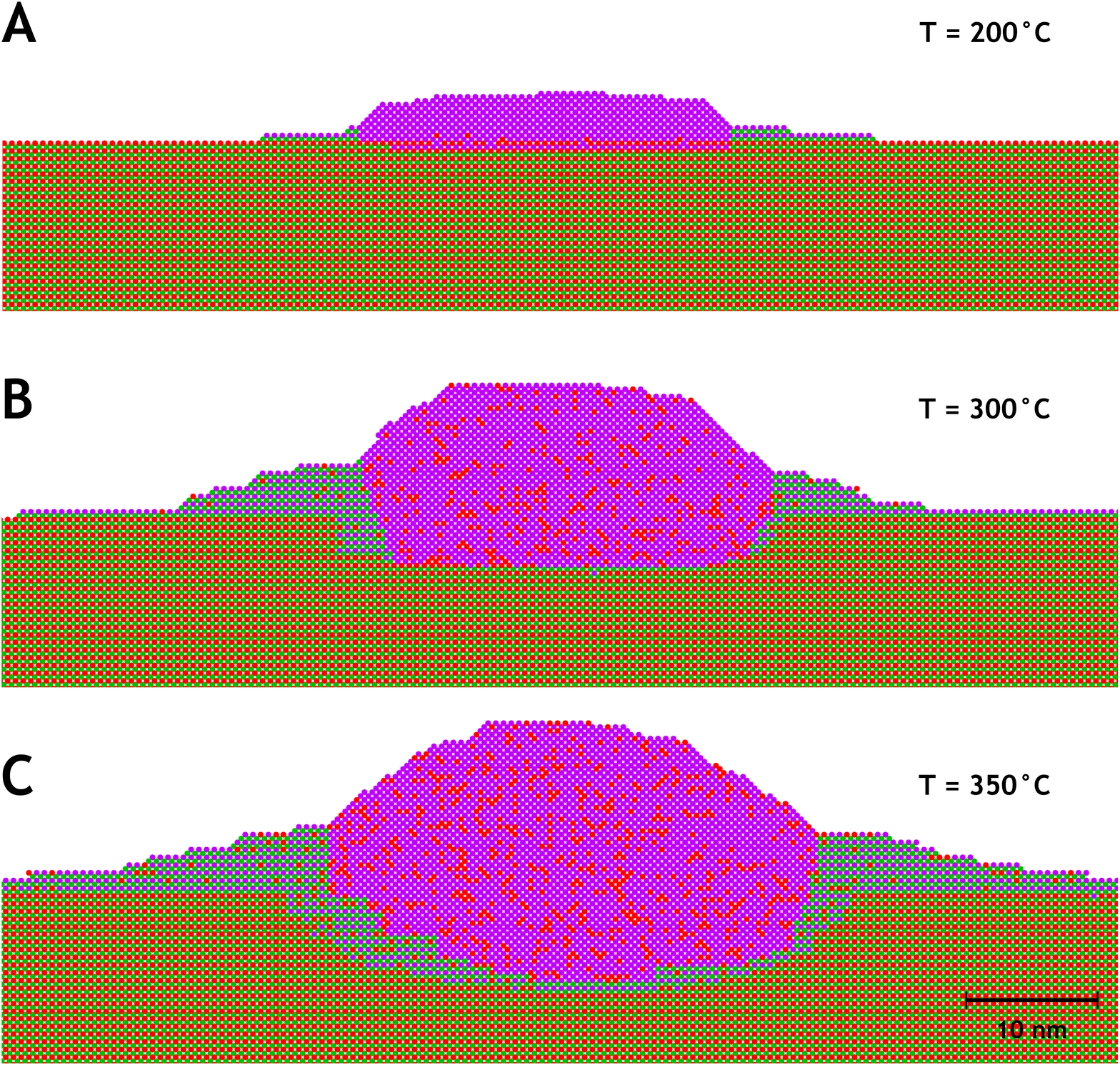} 
\caption{ 
(Color online) Liquid droplets grown at $T=200, 300, \SI{350}{\celsius}$ and $F_{Ga} = 0.1$
ML/s.  Here and throughout the paper, Ga and As atoms initially belonging to
the substrate are colored red and green, respectively. Ga and As atoms deposited
throughout the simulation are colored purple and blue, respectively.  
}
\label{fig_droplet}
\end{figure}

During the first phase of the simulation, Ga atoms are deposited on a flat,
initially As-terminated GaAs substrate. The first monolayer of Ga deposited is
consumed in creating a layer of Ga, resulting in a Ga terminated substrate.
Afterwards, the remaining Ga atoms diffuse along the surface and eventually
nucleate hemispherical droplets as the system attempts to minimize the
vapor-liquid interface. During droplet formation, simulations show that liquid
Ga etches into GaAs, and the amount of etching is regulated by temperature, as
illustrated in Figure \ref{fig_droplet}. Higher temperatures result in more
significant etching.  This is in agreement with experimental observations
\cite{heyn2,wang1}.  As the droplet etches into the substrate liquid Ga atoms
displace substrate As atoms, which subsequently attach near the triple junction.
In addition, some of the displaced substrate material is wicked out of the
droplet in a step-flow growth mode.  The relevant model parameter controlling
the effect of etching is the additional barrier for atom-atom exchanges,
$\epsilon_D$. The value $\epsilon_D = 0.7$ eV was selected to fit qualitative
experimental observations on the amount of etching occurring at various
temperatures.

Once the liquid droplet has formed and come to equilibrium,  an As flux is
introduced to initiate crystallization. Arsenic atoms deposited near or on the
droplet diffuse through the liquid quickly \cite{gorokhov1} and attach most
typically near the triple-junction.  Such crystallization results in a growing
GaAs front and the droplet is crystallized inward. If no nucleation occurs at
the vapor-liquid interface and the GaAs fronts coalesce, a fully crystallized
quantum dot forms in place of the liquid droplet. Such is the case for moderate
temperatures and deposition rate.

\begin{figure*}[!htb]
\centering 
\includegraphics[width=0.8\textwidth]{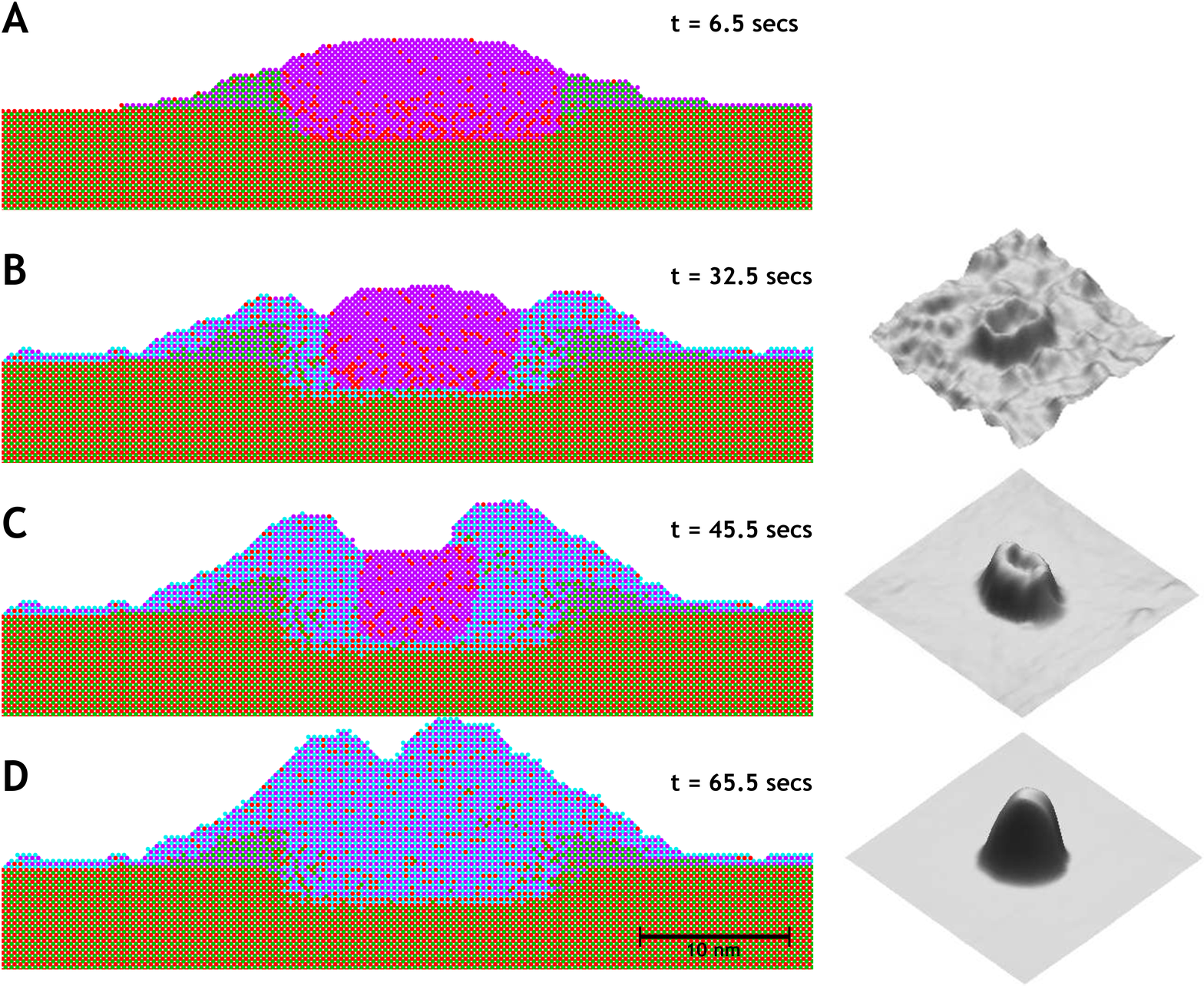} 
\caption{ 
(Color online) Left panel: model snapshots of liquid droplet crystallization at various times
for $T = 275 \SI{}{\celsius}$ and $F_{As} = 0.06$ ML/s resulting in a compact
quantum dot. Ga and As atoms from the original substrate are colored red and
green, respectively.  Ga atoms deposited to form liquid droplets are colored
purple, while As atoms deposited during crystallization are blue. Right panel:
AFM images of the GaAs growth fronts in partially crystallized droplets after 10
seconds (top), 40 seconds (middle) and 90 seconds (bottom). Crystallization was
obtained at $T = 150 \SI{}{\celsius}$ and $5 \times 10^{-7}$ Torr As pressure.
} 
\label{fig_droplet_formation}
\end{figure*}

Figure \ref{fig_droplet_formation} is a sequence of simulation snapshots
illustrating the crystallization of a liquid droplet resulting in a quantum dot
(left panel), along with analogous AFM images of the GaAs fronts obtained
experimentally (right panel).  The simulation images in the figure illustrate a
typical quantum dot grown at $T = 275$\si{\celsius} and $F_{As} = 0.06$ ML/s.
The general trend is that a Ga drop forms once enough Ga has been deposited on
the surface (Figure \ref{fig_droplet_formation}A), followed by crystallization
near the vapor/liquid/solid triple junction upon exposure to As flux (B,C).  As
crystallization progresses, the liquid Ga is consumed, resulting in a fully
crystallized quantum dot (D).  Experimental images (Figure
\ref{fig_droplet_formation}, right panel) were obtained from individual samples
prepared according to the experimental procedure outlined above, varying As
exposure time. The unreacted liquid Ga was removed from the samples before
imaging by selective wet etching \cite{somaschini1}, thus showing GaAs fronts at
various stages during crystallization.  The AFM images confirm the growth
mechanism observed in the simulations during crystallization.

\begin{figure}[!htb]
\centering
\includegraphics[width=\FIGWIDTH]{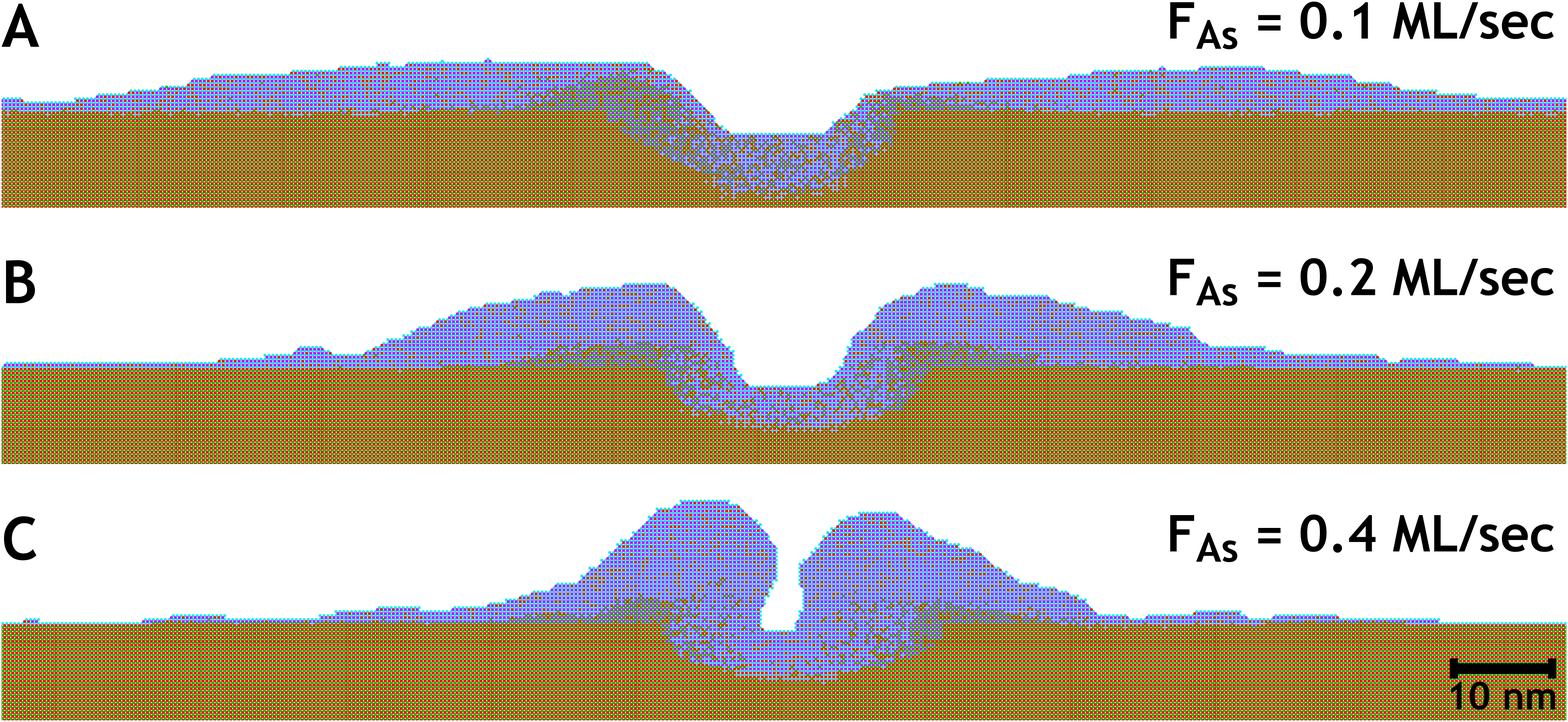}
\caption{
(Color online) Nanorings formed at $T = 375$\si{\celsius} and $F_{As} = 0.10,
0.20$ and $0.40$ ML/s.
} 
\label{fig_nanoring_sim_results}
\end{figure}

Besides compact quantum dots, other nanostructures are predicted by the
simulation by considering a broad range of temperatures and As fluxes.  For
example, if $F_{As}$ is sufficiently low or $T$ is sufficiently large, the
simulations show the formation of nanorings upon crystallization.  The exact
morphology of the nanorings is sensitive to the growth conditions.  Figure
\ref{fig_nanoring_sim_results} shows the morphological dependence of the rings
on $F_{As}$, fixing $T = 375$\si{\celsius}. At low As deposition rate ($F_{As} =
0.1$ ML/s), broad and short nanorings form.  As $F_{As}$ is increased the
nanorings become more compact and taller so that at $F_{As} = 0.4$ ML/s the
resulting structure resembles a ``pitted" quantum dot. This compares well to the
experimental results in Figure \ref{fig_zoology}, e.g. the transition in
structure between Figure \ref{fig_zoology}c and Figure \ref{fig_zoology}f as
As$_4$ BEP is lowered. Figure \ref{fig_zoology}c resembles the pitted quantum
dot structure of Figure \ref{fig_nanoring_sim_results}c. If the BEP is lowered,
the resulting nanostructure in Figure \ref{fig_zoology}f is a broad and shallow
disk surrounding a pit, resembling Figure \ref{fig_nanoring_sim_results}a.
Similar structures and their dependence on both As deposition rate and
temperature have been reported in the literature \cite{somaschini3,
somaschini5}.  

\subsection{Core-Shell Structures} 
In addition to quantum dots and nanorings, simulations show the existence of
Ga/GaAs core-shell structures. These structures consist of liquid Gallium being
completely surrounded by GaAs.  In the low temperature/high deposition rate
regime, the GaAs shells are polycrystalline.  However, in a higher
temperature/lower flux regime, the simulations show the formation of a shell in
registry with substrate. It will be shown below that the first case is the
result of nucleation of GaAs at the vapor-liquid interface whereas the second
case results from a Mullins-Sekerka instability of the crystallization growth
front.

\subsubsection{Nucleation}
\label{sec_nucleation}
\begin{figure}[h]
\centering
\includegraphics[width=\FIGWIDTH]{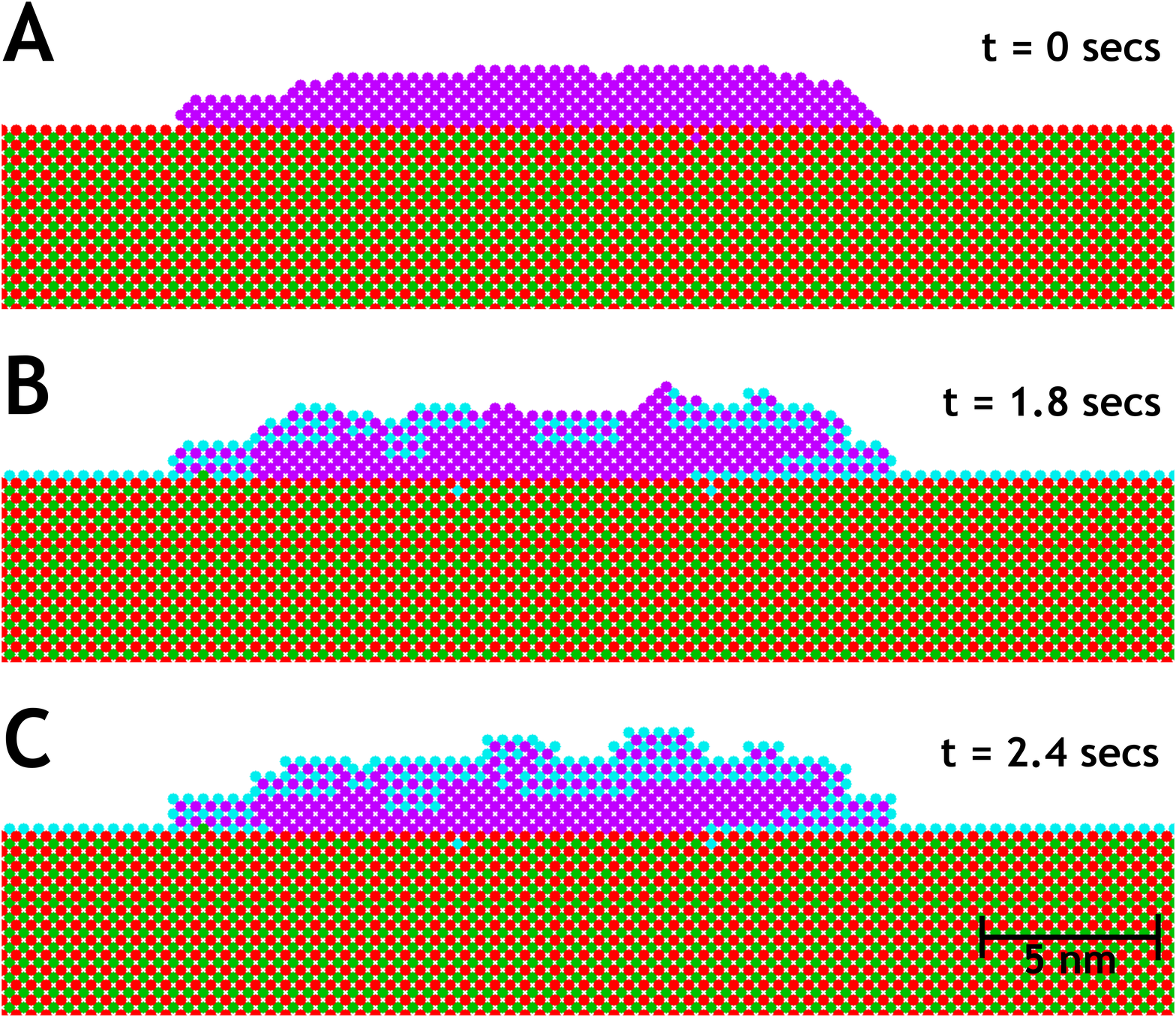}
\caption{
(Color online) Simulation snapshots at times $t = 0, 1.8, 2.4$ seconds after
crystallization, $T = 150$\si{\celsius} and $F_{As} = 0.8$ ML/s.
}   
\label{fig_nucleation}
\end{figure}

In the low temperature and high As deposition rate regime nucleation of GaAs
clusters near the vapor-liquid is significant. This results in the formation of
a polycrystalline GaAs shell surrounding a liquid Ga core, as illustrated in
Figure \ref{fig_nucleation}. In the figure, a liquid droplet grown at $T =
150$\si{\celsius} is crystallized by an As flux, deposited at a rate of 0.8
ML/s.  Nucleation at the vapor-liquid interface occurs within seconds upon
crystallization (Figure \ref{fig_nucleation}B). The liquid core in the final
configuration (C) is completely surrounded by a GaAs shell after 2.4 seconds,
preventing any further crystallization of the liquid. 

\begin{figure}[h] 
\centering 
\includegraphics[width=\FIGWIDTH]{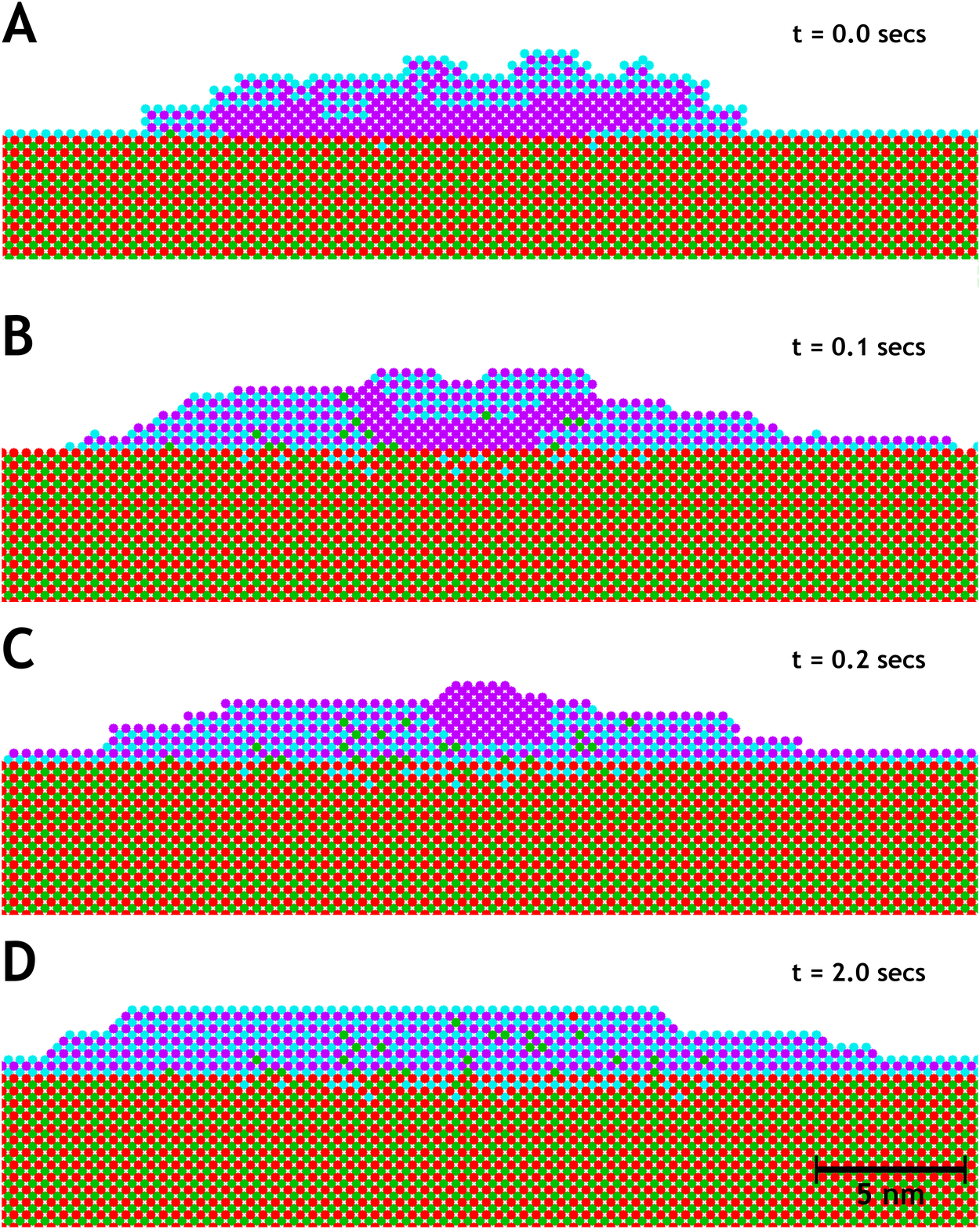}
\caption{
(Color online) Simulation snapshots of a quantum dot annealed at high
temperature at time $t$ after temperature was increased. (A) The dot after
exposure to As deposition at $F_{As} = 0.80$ ML/s and temperature $T =
150$\si{\celsius}. This results in a polycrystalline GaAs shell trapping a
liquid Ga core. (B)-(D) Temperature is increased to $T = 350$\si{\celsius} and
the atoms rearrange in order to fully crystallize the liquid core.
}
\label{fig_anneal} 
\end{figure}

The presence of polycrystalline GaAs, with a high number of grain boundaries and
stacking faults that this implies, prevents the possibility that such islands
could act as efficient emitters and hence are undesirable. Annealing at higher
temperature may remove such defects because grain boundaries and stacking faults
provide fast diffusion paths for the liquid Ga trapped within the shell.  Such
paths are accessible at higher temperatures.  Moreover, thermal fluctuations of
the nuclei can effectively serve to dissolve the shell, which are
characteristically thin in this regime.  Therefore such configurations may be
annealed at a high temperature to remove defects. Indeed, the simulations bear
this out.  Figure \ref{fig_anneal} shows a quantum dot with polycrystalline
shell resulting from the crystallization of a liquid droplet at an As deposition
rate of $F_{As} = 0.80$ ML/s and temperature $T = 150$\si{\celsius} . The dot
is then annealed at a higher temperature $T = 350$\si{\celsius}. The initial
configuration (\ref{fig_anneal}A) shows the droplet prior to recrystallization.
Temperature is then increased, maintaining the same As flux. Ga atoms move along
grain boundaries toward the surface, resulting in a broadening of the dot. The
thin shell dissolves, resulting in liquid Ga exposed to As (B, C). Within two
seconds, the droplet becomes fully crystallized (D) into a shallow GaAs island,
absent of any defects.

These simulations show that liquid cores arising due to nucleation at the
vapor-liquid interface, which would be detrimental to optical and electronic
properties of quantum dots, can be eliminated by annealing.  This expands the
parameter space available for the formation of quantum dots formed via droplet
epitaxy, allowing the fabrication of optical quality dots even at low
temperatures. Even if not strictly related to the presence of a liquid core, but
rather to a low quality of the crystalline nature of the fabricated
nanostructures, post-crystallization \emph{in-situ} annealing temperatures have
been experimentally demonstrated to strongly increase the optical quality of
droplet epitaxy GaAs/AlGaAs dots, reducing individual dot emission
lines\cite{mano2}, allowing single photon emission upto liquid nitrogen
temperatures \cite{cavigli1} and increasing the dot emission decay time.

\subsubsection{Mullins Sekerka Instability}
\label{sec_ms}

As described above, simulations show that crystallization performed at low
temperature and high As flux result in nucleation of polycrystalline GaAs shells
surrounding a liquid Ga core, resulting from nucleation of GaAs at the
vapor-liquid interface. In the high temperature regime, such surface nucleation
does not occur. However, for crystallization at sufficiently high fluxes and
high temperature, simulations  show the existence of liquid Ga core structures
surrounded by GaAs shells in registry with the substrate.  As a consequence,
such shells are monocrystalline and result from a mechanism separate from
surface nucleation. 

\begin{figure}[h]
\centering
\includegraphics[width=\FIGWIDTH]{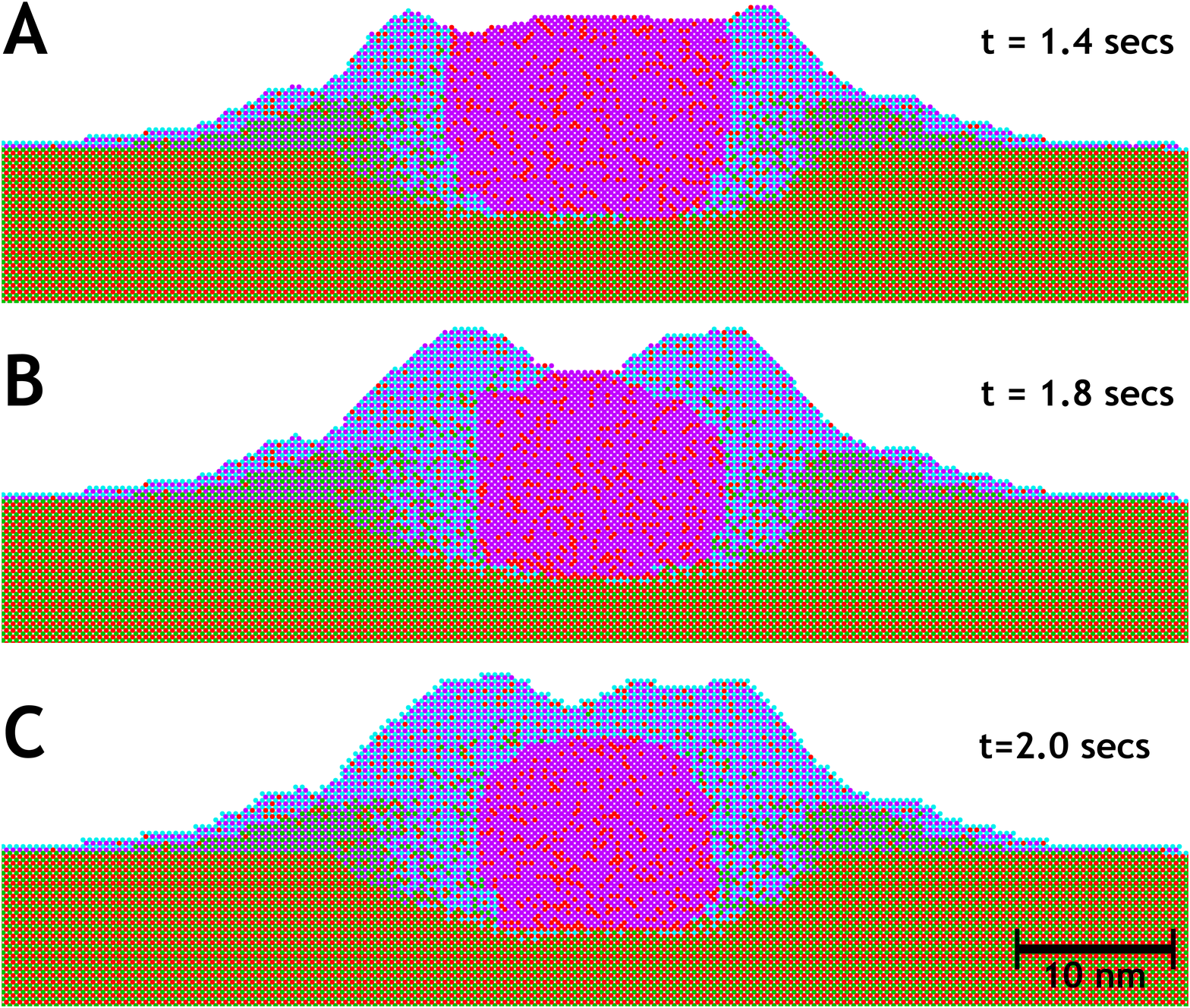}
\caption{ 
(Color online) Snapshots of liquid core formation at times $t = 1.4, 1.8$ and
$2.0$ seconds after crystallization at temperature $T = 350 \SI{}{\celsius}$ and
deposition rate $F_{As} = 1.0$ ML/s.  
}   
\label{fig_ms_instability}
\end{figure}

By examining the formation of such structures in the simulations, we propose
that these shells are driven by an instability at the liquid-solid growth front.
Simulation snapshots in Figure \ref{fig_ms_instability} illustrate the growth
mechanism behind this. When this phenomenon occurs no surface nucleation is
observed;  instead, the growth of the GaAs front undergoes an instability at the
liquid-solid interface characterized by unstable undulations of the solid growth
front. Such instabilities grow along the droplet-vapor surface until they have
completely surrounded the liquid Ga subsequently preventing further
crystallization. 

This behavior suggests the presence of a Mullins-Sekerka (MS) instability during
crystallization, which implies that only perturbations of sufficiently large
wavelength experience unstable growth.  Therefore, droplets must be sufficiently
large to accommodate perturbations of appropriate wavelength in order for the
instability to manifest. The effect of temperature, As flux and droplet radius
have on the presence of the MS instability is given in Section \ref{sec_theory}.
but the droplet size dependence may be utilized in order to confirm the
existence of liquid cores by MS instability experimentally.

It is important to note that simulations suggest that, unlike those formed by
surface nucleation, the core-shell structure formed due to the MS instability
cannot readily be annealed and recrystallized into an epitaxial nanodot. In this
case,  shell is in registry with respect to the substrate.  Therefore, there are
no defects along which liquid Ga can move so that they may crystallize upon
exposure to the As vapor. Crystallization of such liquid cores may still occur,
however, if the shell is thin enough to dissolve upon annealing thus exposing
the liquid Ga to the As flux, however instability-induced shells are
characteristically thicker than those resulting from surface nucleation
according to simulations.

\noindent{\bf Experimental Validation.}
In order to experimentally validate the presence of the MS instability predicted
by simulations, nano-islands were formed via droplet epitaxy by depositing Ga
droplets of extremely different sizes (60 and 250 nm diameter) and crystallizing
them at high As BEP at \SI{150}{\celsius}. Simulations predict a size driven
transition from a uniform crystallization for small droplet radii to a Ga rich
core at higher droplet sizes. We used Si as the growth substrate.  This is
dictated by the need of differentiation of the nano-island constituents (Ga and
As) from that of the substrate (Si) \cite{somaschini6}, thus allowing the use of
standard EDS – STEM techniques for the characterization of the As and Ga
distribution within the islands. The samples were prepared as follows. Before
the introduction into the MBE system, Si(001) substrates were cleaned by
standard RCA treatment and finally dipped into HF solution to get the
H-termination of the surface, confirmed by the ($1\times 1$) pattern observed
with RHEED. Subsequently, the substrate temperature was set at
\SI{780}{\celsius} and the hydrogen desorption was carried out until a mixed
($2\times 1$) ($1\times 2$) surface reconstruction was clearly obtained.  After
this step, the substrate temperature was decreased to either \SI{200}{\celsius}
(sample A) or \SI{600}{\celsius} (sample B) for Ga deposition. Here a Ga
molecular beam flux was supplied with a deposition rate of 0.075 ML/s and a
background pressure below $5\times10^{-10}$ Torr for a total of 3.0 ML of Ga.
Finally the crystallization of Ga was achieved at \SI{150}{\celsius} by exposure
to an As flux of $5\times10^{-5}$ Torr for 5 minutes. During the As irradiation,
the RHEED pattern turned from halo, indicative of liquid droplets, to spotty,
signalling the formation of three dimensional structures. However, a clear
spotty pattern was not observed for the larger islands.

For TEM observations an FEG (Field Emission Gun) TEM/STEM 2200 FS JEOL
instrument operated at 200 kV was used. It was equipped with an energy
dispersive X-ray spectrometer (EDS) and an in-column Omega-type Energy Filter.
By the latter, elemental maps of Ga and As can be obtained in the energy
filtering operation mode of the TEM (EF-TEM). The EF-TEM elemental maps where
acquired with a GATAN slow scan CCD camera controlled by the Digital Micrograph
software. The maps were obtained by recording the intensity of the L3 absorption
edge for Ga (at 1116 eV) and As (at 1323 eV). One post-edge and two pre-edge
images were acquired using a slit of 50 eV for both Ga and As. From the two
pre-edge images the background image was evaluated which was then subtracted
from the post-edge image to get the relevant elemental map. EDS maps were
obtained in the STEM operation mode with a spot size of 1 nm and recording the
intensity of the L characteristic X-ray emission lines of Ga (1096 keV) and As
(1282 keV). The JEOL ultra-thin window Si detector and software were employed
for acquisition of the maps. Dead time during measurements was 5-6\% and
counting time of the order of 8-10 minutes.

\begin{figure}[!hbt] 
\centering
\includegraphics[width=\FIGWIDTH]{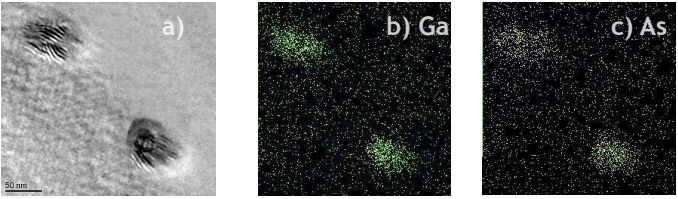}
\caption{
(Color online) Sample A. (a) Planar view TEM micrograph of two (70 nm diameter) GaAs
microislands. (b) and (c) EDS maps of Ga and As, respectively, of the islands
in (a).
}
\label{fig_ms_exp_a}
\end{figure}

\begin{figure}[!hbt]
\centering
\includegraphics[width=\FIGWIDTH]{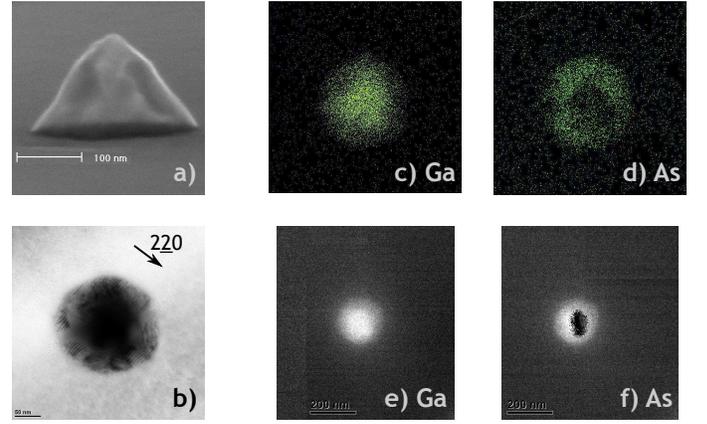}
\caption{
(Color online) Sample B. (a) SEM image of a  GaAs microisland(250 nm diameter). (b) Planar
view TEM micrograph of a GaAs microisland. (c) and (d) EDS maps of Ga and As,
respectively, of the island in (b). (e) and (f) EF-TEM image for Ga and As,
respectively, of another GaAs microisland (TEM image not shown).
}
\label{fig_ms_exp_b}
\end{figure}

The TEM top views of islands from sample A, obtained with diffraction vector $g
= [\text{2-20}]$ clearly show Moir\'{e} fringes, due to the interference between
the crystal lattice of the island (GaAs) and that the Si substrate (Figure
\ref{fig_ms_exp_a}).  EDS map scans show the same distribution of the Ga and As
signals, which mimics the island shape, thus demonstrating the correct
stoichiometry of the GaAs all over the island volume -- i.e. a fully
crystallized island.  A different scenario is shown by sample B.  Here the
nanostructure generated Moir\'{e} interference fringes are visible only at the
island edges with a featureless center.  EDS maps scans show that the
distribution of Ga and As are extremely different inside the island (Figure
\ref{fig_ms_exp_b}). While the Ga signal intensity follows the island profile,
As intensity is peaked at the island perimeter with a small, though non-zero, As
signal coming from the center.  Such a distribution is confirmed by EF-TEM
(energy filtered TEM) measurements, and implies a metallic Ga core similar to
the one depicted in Figure \ref{fig_ms_instability}. 

The experiment rules out the presence of a liquid Ga core/ GaAs shell structure
by means of surface nucleation due to the fact that the shell is monocrystalline
and appears only for the high droplet radius.  Nucleation at the liquid-solid
interface is predicted to be driven more by the growth temperature and As flux.
Because both samples were performed at the same growth conditions, if nucleation
driven GaAs shells were to occur, it must occur in both samples.  Because this
is not the case, we may conclude that the liquid cores observed in the
experiments are driven by an MS instability.

\section{Model Analysis}
\label{sec_theory}

\begin{figure}[!htb]
\centering 
\includegraphics[width=\FIGWIDTH]{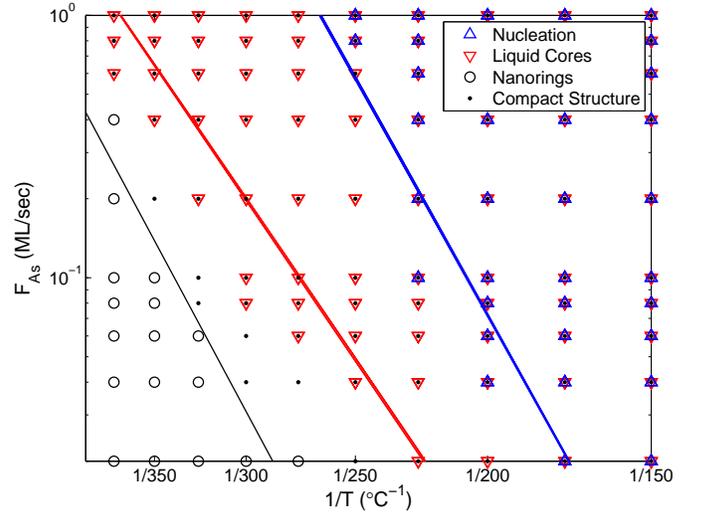}
\caption{
(Color online) Morphological Dependence on Growth Conditions. This
nanostructural phase map summarizes simulation results of droplet epitaxy and
crystallization at various As deposition rates and temperatures. The three
boundary curves indicate theoretically derived critical conditions delineating
the simulation results and obtained in this section. The left-most, black line
is given by Equation {\eqref{eqn_nanoring_line}}. The middle, red curve
corresponds to Equation {\eqref{eqn_liquid_core_line}}. The right-most, blue
curve is given by Equation {\eqref{eqn_nucleation_line}}.
}
\label{fig_morphological_phase_diagram}
\end{figure}

Both simulations and experiments have shown that droplet epitaxy can result in a
wide range of morphologies depending on the growth conditions.  By varying
$F_{As}$ and $T$, we  have established their effect on the resulting
nanostructures observed in simulations.  From this data, it is clear that
compact islands form at low temperature and, as the As overpressure is reduced
or the temperature is increased the nanostructures become rings. In cases of
large $F_{As}$, core-shell structures are observed.  The simulation results are
summarized in the structural map given in Figure
\ref{fig_morphological_phase_diagram}.  In this section, we will appeal to
physical and mathematical arguments to further explain the simulation results.
In particular, the solid lines in Figure \ref{fig_morphological_phase_diagram}
that delineate the morphological structure will be derived in this section.

\begin{figure}[h]
\centering
\includegraphics[width=\FIGWIDTH]{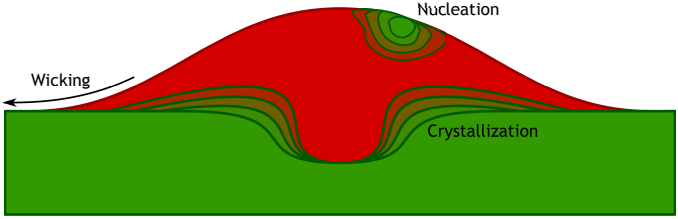}
\caption{
(Color online) Schematic of kinetic processes that determine nanostructural
development.  
} 
\label{fig_droplet_processes}
\end{figure}

We argue that various morphologies observed both in experiments and simulations
can be explained in the context of three key processes active during
nanostructure formation. In the first Ga atoms in the liquid drop will be
``wicked'' out of the droplet onto the substrate by capillary-type forces when
exposed to an As overpressure. These forces arise as it is energetically
favorable for As atoms on the surface to become fully coordinated with Ga atoms.
Next, As atoms deposited near or on the droplet diffuse rapidly \cite{gorokhov1}
through the liquid and attach on a growing GaAs front at the liquid-solid
interface, crystallizing the droplet epitaxially. Finally, As atoms may also
nucleate near the vapor-liquid interface. These three processes: wicking,
crystallization and nucleation are illustrated in Figure
\ref{fig_droplet_processes}. The relative rates at which these processes occur
depend on growth conditions and will determine the resulting morphology.

\subsection{Nanoring Formation}

The formation of nanorings is a competition between the wicking and
crystallization processes. The crystallization process results in GaAs fronts
that grow until the liquid Ga is consumed. If the fronts coalesce before this
occurs, the resulting nanostructure is a compact quantum dot. If however, the
liquid Ga is consumed before the fronts come together, nanorings result.

To compare the rates of the two processes, we establish expressions for the
velocity of the GaAs front as well as the rate at which Ga atoms are wicked out
of the droplet. Assuming quasi-static deposition, the velocity $v_n$ of the GaAs
front under an As deposition rate $F_{As}$ scales according to that rate:
\begin{equation}
    v_n  = g \ell F_{As},
    \label{eqn_normal_velocity}
\end{equation}
where $\ell$ is the atomic spacing of the lattice and $g$ is factor that depends
shape of the liquid region and possible difference in the As adsorption
probability between the droplet itself and its surroundings. For simplicity,
$g=1$. In the wicking process, Ga atoms are driven from the droplet by the
deposition of As on the surface away from the droplet. The time scale of the
wicking process is given by
\begin{equation}
    \tau = \frac{1}{F_{As}},
    \label{eqn_wicking_timescale}
\end{equation}
while the diffusion length scale is given by
\begin{equation}
    \lambda = \sqrt{\frac{D^\prime_{Ga}}{F_{As}}},
    \label{eqn_wicking_lengthscale}
\end{equation}
where $D^\prime_{Ga}$ is the diffusion coefficient of Ga diffusing on a mostly
Ga terminated surface, in the presence of As deposition. We have 
$$ D^\prime_{Ga} = \ell^2 R_0\exp\left[-\frac{E^\prime_{Ga}}{k_B T}\right],$$
where $R_0 = 10^{13} \text{s}^{-1}$ was defined in Section \ref{sec_model} to be a
rate prefactor.

The energy barrier $E^\prime_{Ga} = 1.26$ eV describing the diffusion of Ga on
surface not purely Ga terminated is obtained from the harmonic average of the
diffusion rates of Ga-on-Ga diffusion and Ga-on-As diffusion. By prescribing a
constant value $E^\prime_{Ga}$ that describes the effective energetic barrier
for diffusion on a mostly Ga-terminated surface, we have assumed that the
diffusivity throughout the domain is constant. In reality, diffusivity is
spatially dependent due to differences in surface reconstruction near and away
from the liquid Ga as observed in micro-RHEED experiments by Isu et al.
{\cite{isu1}}.  This effect is also at the origin of the possibility, offered by
droplet epitaxy, to tune ring morphologies {\cite{somaschini6}}. In extreme
growth regimes when temperature is less than $150^\circ$C and As overpressure
greater than $5\times 10^{-5}$ Torr, the surface reconstruction may change
rapidly even in the proximity of the droplet, implying that the spatial
dependence on diffusivity is an important effect. Away from such growth
conditions, however, assuming constant effective diffusivity is a reasonable
approximation.

Equations \eqref{eqn_wicking_timescale} and \eqref{eqn_wicking_lengthscale}
yield a velocity $v_w$ for the wicking of Ga atom from the liquid droplet
\begin{equation}
    \label{eqn_wicking_velocity}
    v_{w} = \frac{\lambda}{\tau} = \ell\sqrt{R_0 F_{As}}\exp\left[-\frac{E^\prime_{Ga}}{2 k_B T}\right].
\end{equation}
A fully crystallized quantum dot forms under conditions where the
crystallization process is dominant ($v_n \gg v_w$). However, if the wicking
process is sufficiently fast ($v_n \sim v_w$), then as described above, the
fronts may fail to coalesce, resulting in a nanoring. The critical configuration
separating these two scenarios occurs when the crystallization fronts are
tangent to each other. To express this critical condition quantitatively,
consider the volume of unconsumed liquid Ga at time $t$: 
\begin{equation}
    V(t) = V_0 - 2 \ell v_w t - 2\rho_1(v_nt)^2,
    \label{eqn_droplet_volume}
\end{equation}
where $\rho_1 = \frac{\pi}{4}$ is a shape constant describing the geometry of
the crystallization front (which we model as a quarter-circle) and 
$$V_0 = \rho_2 r^2,$$ 
is the initial volume of the liquid droplet of radius $r$, $\rho_2 =
\frac{3\pi}{4}$ being the shape constant that describes the geometry of the
droplet. As a first-order approximation, the value of $\rho_2$ is selected as
the average of the shape constant describing a circular and half-circular
droplet. The radius of a Ga droplet obtained from the deposition of 4.0 ML of Ga
atoms at temperature $T$ and flux $F_{Ga}$ was found empirically from
independent simulations to follow the power law:
\begin{equation}
     r = \ell^{1-2\alpha}r_0 \left(\frac{D_{Ga}}{F_{Ga}}\right)^\alpha
     \label{eqn_droplet_size},
\end{equation}
where $r_0 = 11.34$ (\# atoms) and $\alpha = 0.182$ are empirically obtained
values.  $D_{Ga}$ is the diffusion coefficient for Ga-on-Ga diffusion with
corresponding energy barrier $E_{Ga} = 0.9$ eV. The positive root $t_f$ of
equation \eqref{eqn_droplet_volume} describes the time when all Ga has been
consumed. The length of the crystallized front is then given by $v_n t_f$, and
the critical condition may be expressed as 
$$v_n t_f = r$$
Using equations \eqref{eqn_droplet_size}, \eqref{eqn_normal_velocity}, and
\eqref{eqn_wicking_velocity} , this critical configuration can be written in
terms of the growth conditions as
\begin{widetext}
\begin{equation}
    \label{eqn_nanoring_line}
    F_{As}^r =  \left(\frac{2F_{Ga}^\alpha}{\left(\rho_2 - 2\rho_1\right)gr_0}\right)^2 
        R_0^{1-2\alpha} 
        \exp\left[\frac{2\alpha E_{Ga} - E^\prime_{Ga}}{k_B T}\right].
\end{equation}
\end{widetext}
If $F_{As} >  F_{As}^r$ then compact structures will result on other hand if
$F_{As} < F_{As}^r$ nanorings will result. Figure \ref{fig_nanoring_sim_results}
illustrates the transition from nanorings to compact quantum dots with
increasing $F_{As}$.  The black line in Figure
\ref{fig_morphological_phase_diagram} shows a plot of $F_{As}^r$ vs. $1/T$.  The
agreement with the simulation data is quite good -- it separates compact
structures (black dots in the figure) from nanorings (black circles).

A crucial assumption has been made in the argument presented here, namely that
no nucleation takes place within the Ga droplet and the crystallizing front
moves in a stable fashion (i.e. no Mullins-Sekerka instability occurs).  In what
follows we shall examine in what regimes in parameter space these factors play
an important role.

\subsection{Nucleation}

Here we will demonstrate that when the As deposition rate is sufficiently large
and temperature is low enough, the dominant process will be nucleation at the
vapor-liquid interface. As a consequence, the wicking process makes a negligible
contribution to the morphology. The rate of the nucleation process at the
vapor-liquid interface may be estimated by considering the As concentration
$c_{As}(x,y,t)$ within a domain of liquid Ga in contact with a GaAs substrate
and As flux. For simplicity we will consider the domain to be rectangular with
periodic boundary conditions in the horizontal direction. Assuming quasi-static
deposition at a rate $F_{As}$ ML/s and temperature $T$, the concentration
satisfies 
\begin{eqnarray*}
    \nabla^2 c_{As} &=& 0, \\
    \ell D_{As} \frac{\partial c_{As}}{\partial y}\bigg|_{y = L} &=& F_{As}, \\
    c_{As}\bigg|_{y = 0} &=& c_0
\end{eqnarray*}
where $c_0$ is the equilibrium As concentration above a flat liquid-solid
interface and the diffusion coefficient 
$$ D_{As} = \ell^2 R_0\exp\left[-\frac{\epsilon_D}{k_B T}\right], $$ 
describes the diffusion of As through liquid Ga,  $\epsilon_D = 0.7$ eV being
the energy barrier for diffusion throughout the liquid droplet defined earlier
in Section \ref{sec_model}.

The equilibrium concentration $c_0$ is of the form 
$$ c_0 = \ell^{-2}\exp\left[\frac{E_A - E_D}{k_B T}\right],$$
where $E_D, E_A$ are the energy barriers for the detachment and attachment of an
As atom in the liquid phase from and onto a flat interface, respectively,
defined in Section \ref{sec_model}.  From this diffusion model, the equilibrium
concentration $c_{surf}$ at the surface of the droplet is given by
\begin{equation}
\label{eqn_surface_concentration}
c_{surf} = \frac{F_{As} L}{\ell^3 R_0} \exp\left[\frac{\epsilon_D}{k_B T}\right] + 
    \ell^{-2}\exp\left[\frac{E_A - E_D}{k_B T}\right],
\end{equation}
where $L$ is the height of the droplet.

Nucleation is most likely to occur where the As concentration is the largest,
near the vapor-liquid interface. This means that nucleation will occur when
$c_{surf}$ is larger than some critical concentration: 
\begin{equation}
    c_{surf} \geq c_0 \exp\left[\frac{E_{nuc}}{k_B T}\right],
    \label{eqn_nucleation_crit}
\end{equation}
where $E_{nuc} = 0.01$ eV is the nucleation barrier of GaAs in liquid Ga,
treated as a fitting parameter. Replacing $L$ in equation
\eqref{eqn_surface_concentration} with the droplet radius $r$, the critical
condition \eqref{eqn_nucleation_crit} for surface nucleation can be expressed in
terms of $T$ and $F_{As}$ as: 
\begin{widetext}
\begin{equation}
    \label{eqn_nucleation_line}
    F_{As}^n = \frac{R_0^{1-\alpha}F_{Ga}^\alpha}{r_0} 
        \exp\left[\frac{E_A - E_D - \epsilon_D + \alpha E_{Ga}}{k_B T}\right]
        \left(\exp\left[\frac{E_{nuc}}{k_B T}\right] - 1 \right).
\end{equation}
\end{widetext}
If the As deposition rate exceeds $F_{As}^n$ then GaAs crystallites will form at
the vapor-liquid interface, as illustrated in Figure \ref{fig_nucleation}. The
blue line in Figure \ref{fig_morphological_phase_diagram} is a plot $F_{As}^n$
as function of $1/T$. It accurately predicts the presence of nucleation in
simulation results (blue triangles).

\subsection{Mullins-Sekerka Instability}

In the case where the rates of both the wicking and nucleation processes are
negligible, the crystallization process dominates, resulting in the growth of
crystallization fronts at the triple point and in registry with the substrate.

As observed in Section \ref{sec_ms}, this growth can be unstable due to a
Mullins-Sekerka instability leading to GaAs shells epitaxial to the substrate
surrounding a liquid Ga core.  Perturbations to the planar growth front of
sufficiently long wavelength experience this instability, and a standard linear
perturbation analysis \cite{bales1, caroli1} yields a critical wavelength
$$ \Lambda_c = 2\pi\ell^{3/2}\sqrt{\frac{R_0\gamma}{F_{As}k_B T}} \exp\left[\frac{E_A-E_D-\epsilon_D}{2 k_B T} \right],$$
where $\gamma = 0.1$ eV is the liquid/solid interfacial energy, obtained
directly from the model. In order to accommodate perturbations that experience
the MS instability, droplets necessarily must have radius on the order of
$\Lambda_c$, i.e. $r \geq C \Lambda_c$, for some constant $C$.  This critical
condition may be expressed  in terms of $F_{As}$ and $T$ using the above
equation along with the model for droplet radius \eqref{eqn_droplet_size}:
\begin{widetext}
\begin{equation}
    \label{eqn_liquid_core_line}
    F_{As}^{ms} = C^2 \ell\left(\frac{2\pi F_{Ga}^\alpha \ell}{r_0}\right)^2 \frac{\gamma}{k_B T} R_0^{1-2\alpha} 
             \exp\left[\frac{E_A - E_D - \epsilon_D + 2\alpha E_{Ga}}{k_B T}\right].
\end{equation}
\end{widetext}
If $F_{As}^{ms} < F_{As} < F_{As}^n$  then liquid cores will form via a
Mullins-Sekerka instability.  Figure \ref{fig_morphological_phase_diagram} shows
a plot of $F_{As}^{ms}$ vs $1/T$. The scaling constant $C = \frac{1}{8}$ was
selected to best match simulation results, though its specific value do not
affect the qualitative shape of the boundary curve, and in particular its slope.
The simulation results when a liquid core was observed are plotted as red
triangles. We observe that the theoretical curve slightly underestimates the
instability within the simulations. This underestimation is inherent to our
model and can be attributed to discrete effects. Such effects on nucleation and
hence instabilities of the type outlined above are indeed well-studied in the
context of diffusion limited aggregation \cite{russo1, witten1}.

\section{Conclusion}

We have presented experimental, simulation and analytical results detailing the
precise relationship between growth conditions and the resulting nanostructures
formed in droplet epitaxy and crystallization experiments.  The KMC model used
in conjunction with experiments was presented as an explicitly atomistic,
multi-species, multi-phase model capable of simulating all the relevant
processes involved with GaAs homoepitaxy, droplet epitaxy and crystallization.
As an initial validation, the KMC model accurately reproduces surface
termination diagram in the case of GaAs homoepitaxy. We then presented
simulation and experimental results of droplet formation and crystallization,
exhibiting a qualitative agreement on the resulting morphological dependence on
As flux and temperature. Both experiments and simulation suggest a continuum of
structures ranging between compact quantum dots to broad nanorings as a function
of As flux and temperature. Simulations also suggest the existence of Ga/GaAs
core shell structures, as well as elucidating the mechanisms behind their
formation. The simulations suggested that a quantum dot with a liquid core could
result due to a Mullins-Sekerka instability during the growth of the GaAs front.
We presented experiments which validate this hypothesis. In the case of
nucleation-driven shells, we provide simulation evidence suggesting that under
high temperature annealing, polycrystalline shells may be recrystallized into an
epitaxial GaAs island. Lastly, we developed a unifying theory identifying three
key processes active during crystallization. It predicts the existence of all
the phenomena observed above and their dependence on growth conditions and
compares well with empirical simulation data.

\section{Acknowledgements} 

P.S. and K.R. were supported in part by NSF support grants DMS-0854870 and
DMS-1115252. D.N. and J.M.M. acknowledge the financial support of the NSF
through Grant No. DMR-0906909.  S.B., C.S. and S.S. acknowledge the CARIPLO
Foundation for financial support (prj. SOQQUADRO - no. 2011-0362)

\bibliography{mybib}{}
\bibliographystyle{apsrev}

\end{document}